\documentclass[amsmath,superscriptaddress,floatfix,twocolumn]{revtex4}
\usepackage{graphicx}
\usepackage{amssymb}
\begin{document}

\title{Spin Glass Order Induced by Dynamic Frustration}
\author{E. A. Goremychkin}
\affiliation{Materials Science Division, Argonne National Laboratory, Argonne, IL 60439-4845, USA}
\affiliation{ISIS Pulsed Neutron and Muon Facility, Rutherford Appleton Laboratory, Chilton, Didcot, Oxfordshire, OX11 0QX, United Kingdom}
\author{R. Osborn}
\affiliation{Materials Science Division, Argonne National Laboratory, Argonne, IL 60439-4845, USA}
\author{B. D. Rainford}
\affiliation{Department of Physics and Astronomy, University of Southampton, Southampton, SO17 1BJ, United Kingdom}
\author{R. T. Macaluso}
\affiliation{Program of Chemistry and Biochemistry, University of Northern Colorado, Greeley, CO 80639, USA}
\author{D. T. Adroja}
\affiliation{ISIS Pulsed Neutron and Muon Facility, Rutherford Appleton Laboratory, Chilton, Didcot, Oxfordshire, OX11 0QX, United Kingdom}
\author{M. Koza}
\affiliation{Institut Laue Langevin, F-38042 Grenoble C{\'e}dex, France}

\date{\today}

\begin{abstract}
Spin glasses are systems whose magnetic moments freeze at low temperature into random orientations without long-range order. It is generally accepted that both frustration and disorder are essential ingredients in all spin glasses,  so it was surprising that PrAu$_2$Si$_2$, a stoichiometric compound with a well-ordered crystal structure, was reported to exhibit spin glass freezing.  In this article, we report on inelastic neutron scattering measurements of the crystal field excitations, which show that PrAu$_2$Si$_2$ has a singlet ground state and that the exchange coupling is very close to the critical value to induce magnetic order.  We propose that spin glass freezing results from dynamic fluctuations of the crystal field levels that destabilize the induced moments and frustrate the development of long-range magnetic correlations.  This novel mechanism for producing a frustrated ground state could provide a method of testing the concept of `avoided criticality' in glassy systems.
\end{abstract}

\maketitle

Frustration arises from competing interactions that favour incompatible ground states \cite{Fischer 1991, Ramirez 2003}.  For example, in rare earth intermetallic compounds, the magnetic moments of the \textit{f}-electrons on each rare earth site interact with neighbouring moments through RKKY exchange interactions that oscillate in sign with increasing separation.  Except for specific classes of geometrically frustrated lattices, well-ordered crystal structures produce either ferromagnetic or antiferromagnetic order depending on the energy minimization of the interactions between all the neighbouring moments.  However, when there is disorder, either in the site occupations or the exchange interactions, the additional randomness can prevent a unique ordered ground state.  Instead, these systems may form spin glasses, which possess a multitude of possible disordered ground states, into one of which the system freezes below the glass transition temperature, T$_g$ \cite{Fischer 1991}.  In the thermodynamic limit, spin glasses display broken ergodicity, preventing significant fluctuations to any of the other degenerate spin configurations.  

In recent years, two stoichiometric intermetallic compounds have displayed evidence of spin glass order, URh$_2$Ge$_2$ \cite{Sullow 1997} and PrAu$_2$Si$_2$ \cite{Krimmel 1999}, both nominally with the same tetragonal (ThCr$_2$Si$_2$-type) crystal structure.  In both samples, classic spin glass behaviour was observed (T$_g$ = 11~K and 3~K, respectively), with a frequency dependent peak in the \textit{ac} susceptibility and irreversibility in the field-cooled and zero-field-cooled magnetizations.  It was quickly established that spin-glass freezing in the uranium compound resulted from site disorder on the rhodium and germanium sublattices.  Extended annealing to remove this source of disorder was sufficient to transform samples into ordered antiferromagnets \cite{Sullow 2000}.  On the other hand, the spin glass behaviour of PrAu$_2$Si$_2$ is very robust, and appears in the best quality samples after extensive annealing.  A recent M\"{o}ssbauer study, combined with neutron and x-ray diffraction, concluded that interchange of gold and silicon atoms was less than 1\% \cite{Ryan 2005}.  Furthermore,  the intentional introduction of disorder through the substitution of germanium for silicon \textit{stabilizes} long-range antiferromagnetic order at concentrations greater than 12\% \cite{Krimmel 1999b}.  Finally, the praseodymium sublattice is face-centred tetragonal, and does not contains any of the triangular or tetrahedral motifs normally associated with geometric frustration \cite{Ramirez 2003}.  This article seeks to explain the origin of magnetic frustration in PrAu$_2$Si$_2$, given that conventional models involving static disorder or lattice topology do not seem to apply.

Firstly, it is essential to establish the nature of the praseodymium \textit{f}-electron magnetism. The nine-fold degeneracy of the 4$f^2$-electron states is lifted by the crystal field potential of tetragonal symmetry produced by the surrounding gold and silicon ions giving a set of singlet and doublet levels.  The doublet states have finite magnetic moments, but the singlet states are non-magnetic, except in the presence of exchange interactions with neighbouring sites as discussed below.  Transitions between the crystal field levels can be observed directly by neutron scattering as inelastic peaks, whose energies and intensities can be used to refine the parameters of the crystal field potential.  The peak widths are inversely proportional to the lifetimes of the excited states, which, in metals, are limited by conduction electron scattering.

We have used neutron scattering to study samples of PrAu$_2$(Si$_{1-x}$Ge$_x$)$_2$, with $x$ = 0, 0.05, 0.10, 0.15, 0.20, 0.80, and 1, which were synthesized by arc melting followed by annealing in vacuum for four weeks at 850$^\circ$C. Neutron diffraction measurements on PrAu$_2$Si$_2$ showed no evidence of any additional phases and measurements of the  \textit{dc} susceptibility in a field of 20~Oe showed the previously observed difference between field-cooled and zero-field-cooled susceptibilities confirming a spin glass freezing temperature of 3~K. The inelastic neutron scattering measurements were carried out at the Institut Laue-Langevin (Grenoble, France) on the high resolution time-of-flight spectrometer IN6, using an incident energy of 3.1~meV, at temperatures ranging from 1.5~K to 30~K.  

\begin{figure}[tb]
\vspace{-0.2in}
\includegraphics[width=2.5in]{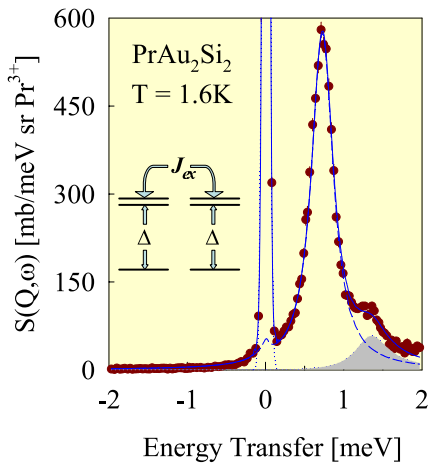} 
\vspace{-0.2in}
\caption{\textbf{Crystal field transition in PrAu$_2$Si$_2$}. Inelastic neutron scattering from PrAu$_2$Si$_2$ measured at 1.5~K on IN6 with an incident energy of 3.1~meV. The solid line is a fit to the singlet-doublet crystal field transition at $\Delta=0.7$ meV (dashed line) and an elastic resolution-limited peak from nuclear incoherent scattering.  The shaded area represents double-scattering from the transition at $2\Delta$. Further details of the crystal field refinement are given in Ref. \cite{Goremychkin 2007}. The inset illustrates the mechanism for induced moment formation, in which interionic exchange coupling, $J_{ex}$, admixes the excited magnetic doublet into the singlet ground state.
 \label{Fig1}}
\end{figure}

In an earlier report, we determined the phenomenological parameters of the crystal field potential in PrAu$_2$Si$_2$ \cite{Goremychkin 2007} and showed that the ground state is a non-magnetic singlet with the first excited level a magnetic doublet at an energy $\Delta=0.7$ meV (see Fig. 1).  The remaining levels are all above 7~meV, so this is effectively a two-level system at temperatures close to the glass transition.  

\begin{figure}[tb]
\vspace{-0.2in}
\includegraphics[width=3in]{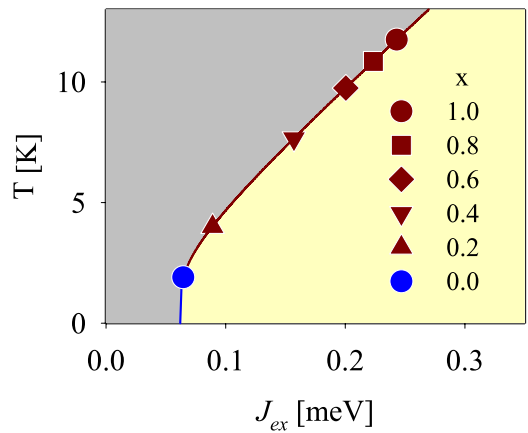} 
\vspace{-0.2in}
\caption{\textbf{Phase diagram of PrAu$_2$(Si$_{1-x}$Ge$_x$)$_2$}. A self-consistent mean-field calculation of the phase diagram of PrAu$_2$(Si$_{1-x}$Ge$_x$)$_2$ as a function of the interionic exchange energy, $J_{ex}$ for a singlet-doublet transition energy, $\Delta=0.7$~meV. The light area represents the region in which spontaneous magnetic moments are induced by the exchange.  The red circles are the antiferromagnetic transition temperatures as a function of germanium concentration, $x$.  The blue circle is the spin glass freezing temperature in PrAu$_2$Si$_2$.
 \label{Fig2}}
\end{figure}

Before discussing the origin of spin glass behaviour, we review what is known about magnetic order in singlet ground-state systems.  Since the ground state of the isolated ion is non-magnetic, a magnetic moment can only result from an admixture of excited crystal field states produced by exchange interactions with neighbouring praseodymium ions.  There is a considerable body of research on such induced moment systems, both theoretical \cite{Grover 1965, Cooper 1972} and experimental \cite{Birgeneau 1971, Houmann 1975, Blanco 1997}, showing that, for any given value of the low-lying crystal field transition energy ($\Delta$), there is a critical value of the exchange energy ($J_{ex}$). Below this critical value, the system remains paramagnetic (\textit{i.e.}, a Van Vleck paramagnet) at all temperatures.  However, when the exchange is sufficiently strong, there is a critical temperature below which a ground state moment forms spontaneously.  In a two-level system, this temperature is given in a mean field model by
\begin{equation}
T_c = \Delta  \bigg\{\ln\left[{\frac{J_{ex}\alpha^2+n\Delta}{J_{ex}\alpha^2-\Delta}}\right] \bigg\} ^{-1}
\end{equation}
where $\alpha$ is the dipole matrix element coupling the two levels and $n$ is the degeneracy of the excited state.  A calculation of $T_c$ as a function of $J_{ex}$ for $\Delta=0.7$~meV and $n=2$ is shown in Fig. 2.

In most induced moment systems, $T_c$ marks the transition to long-range order.  Examples include Pr \cite{Houmann 1975}, Pr$_3$Tl \cite{Birgeneau 1971}, Pr$_3$In \cite{Heiniger 1975} and PrNi$_2$Si$_2$ \cite{Blanco 1997}.  However, in the presence of static disorder, the low-temperature phase could also be a spin glass, as shown by Sherrington \cite{Sherrington 1979} in a mean field model developed to explain scandium-terbium alloys \cite{Sarkissian 1976} and PrP$_{0.9}$ \cite{Yoshizawa 1983}.  Sherrington calculated that a sufficiently large distribution of exchange interactions, ${\delta}J_{ex}$, could lead to spin glass freezing.  There is a multicritical point at 
\begin{equation}
zJ_{ex} = \sqrt{z}\;\delta\!J_{ex}=\Delta/(2\alpha^2)
\end{equation}
where $z$ is the number of nearest neighbours.  At larger values of $\delta\!J_{ex}$, spin glass order would supercede long-range order. In the case of PrAu$_2$Si$_2$, for which $z=4$, $\delta\!J_{ex}$ should be more than twice $J_{ex}$ so Sherrington's model requires a high degree of disorder to generate spin glass behaviour.

\begin{figure*}[tb]
\vspace{-0.2in}
\includegraphics[width=6.5in]{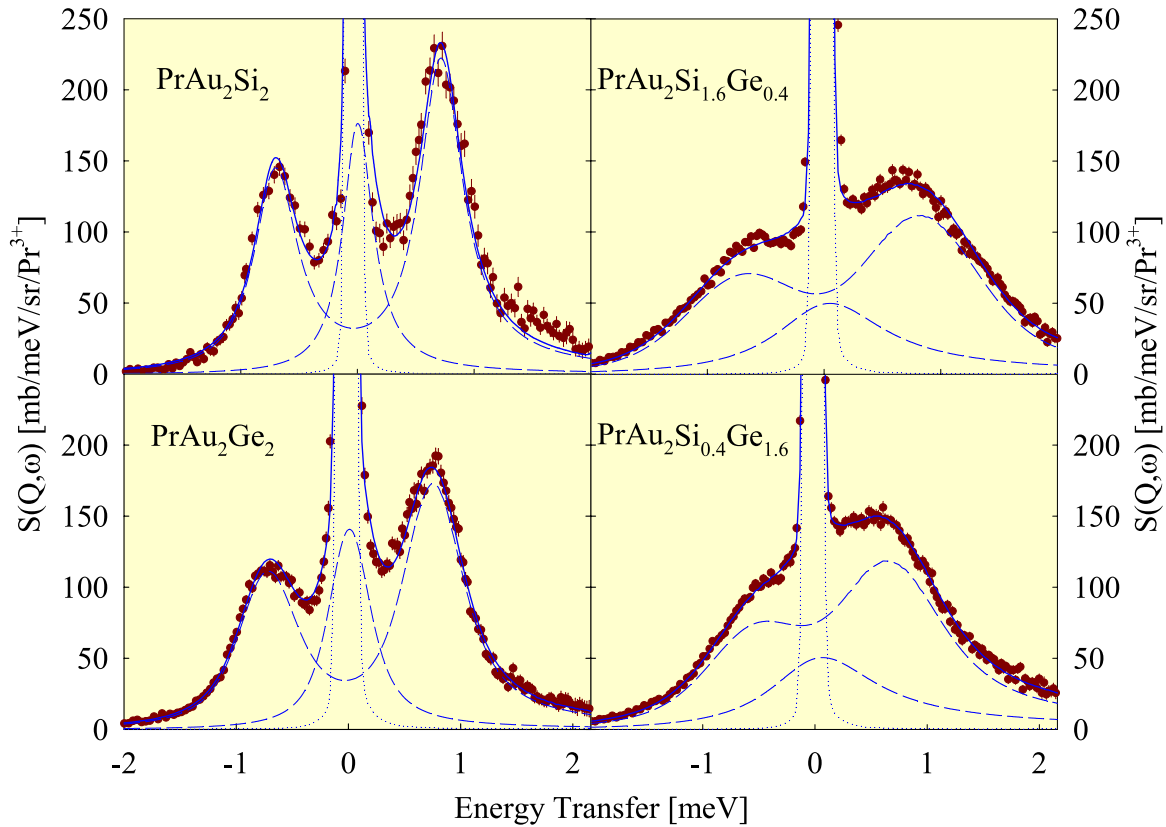} 
\vspace{-0.2in}
\caption{\textbf{Crystal field transitions in PrAu$_2$(Si$_{1-x}$Ge$_x$)$_2$}. Inelastic neutron scattering from the singlet-doublet transition as a function of germanium doping at 20K (red circles).  The solid line is a fit to two components shown as dashed lines; the inelastic singlet-doublet transition and quasielastic scattering within the excited doublet.  The fitted peaks have been broadened by a convolution of Lorenzian and Gaussian lineshapes with the instrumental resolution.
\label{Fig3}}
\end{figure*}

Although such strong disorder appears unlikely in PrAu$_2$Si$_2$, we have explored the Sherrington model further by deliberately disordering the silicon sublattice through germanium substitution.  Krimmel \textit{et al} had already found that long-range antiferromagnetism in PrAu$_2$(Si$_{1-x}$Ge$_x$)$_2$ is stabilized at concentrations $x>0.12$ \cite{Krimmel 1999b}.  Our inelastic neutron scattering measurements shows that the energy of the singlet-doublet transition is nearly independent of $x$ (Fig. 3).  What does change is the strength of the interionic exchange, $J_{ex}$.  In Figure 2, we estimate the strength of $J_{ex}$ \textit{vs} $x$ subsituting the measured values of $T_c$ in equation 1.  This shows that $J_{ex}$ increases by a factor 4 from $x=0$ to $x=1$.  Germanium doping should therefore be an ideal way of enhancing the exchange disorder that is central to the Sherrington model.

\begin{figure}[b]
\vspace{-0.2in}
\includegraphics[width=3.5in]{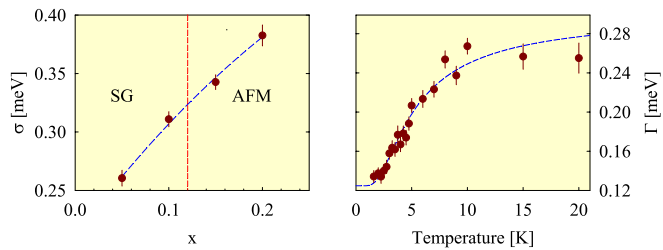} 
\vspace{-0.2in}
\caption{\textbf{Broadening of the singlet-doublet crystal field transition}.  (a)  The standard deviation of the Gaussian broadening of the 0.7~meV transition as a function of germanium doping, $x$, measured at 20~K, \textit{i.e.}, above $T_c$.  The red dashed line marks the critical concentration separating spin glass from antiferromagnetic order. (b) The Lorenzian half-width of the 0.7~meV transition in PrAu$_2$Si$_2$ as a function of temperature.  The blue dashed lines are guides to the eye.
 \label{Fig4}}
\end{figure}

The most noticeable effect of chemical disorder is the large increase in the energy width of the singlet-doublet transition (Fig. 3).  There are two main contributions to the overall width; lifetime broadening, which is Lorenzian in shape and will be discussed later, and inhomogeneous broadening from the chemical disorder, which has a Gaussian form \cite{de Gennes 1958}.  If we convolve both contributions with the instrumental resolution to produce the fits in Fig. 3, we estimate that the Gaussian width rises approximately linearly with $x$, for $x\leq0.2$ (Fig. 4a).  This shows that there is no correlation between the measured inhomogeneous broadening and the transition to long-range antiferromagnetic order and confirms that static disorder is not the critical factor determining the occurrence of spin glass freezing.

\begin{figure*}[tbp]
\vspace{-0.2in}
\includegraphics[width=6.5in]{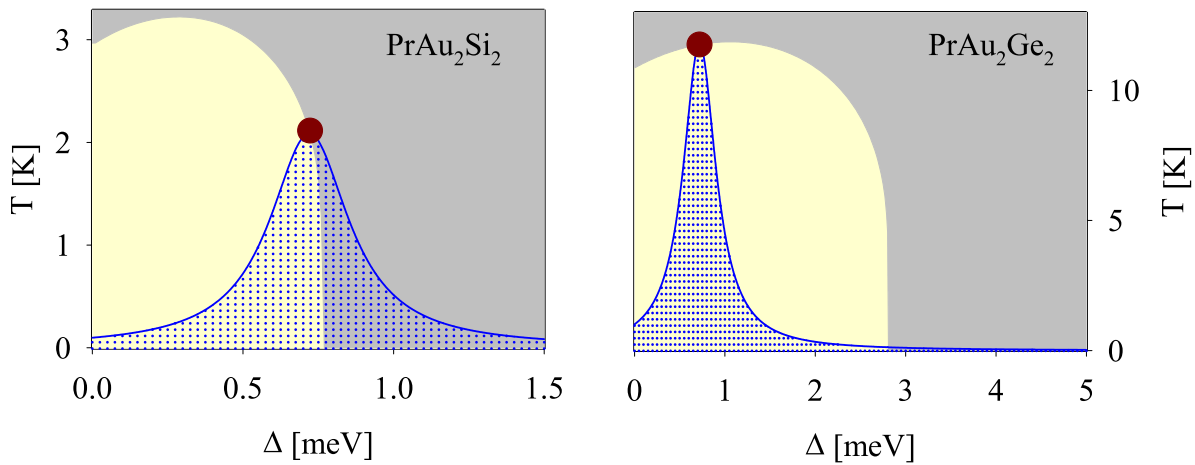} 
\vspace{-0.2in}
\caption{\textbf{Dynamic instability of induced moment formation}.  The mean field calculation of the phase diagram of PrAu$_2$Si$_2$ and PrAu$_2$Ge$_2$ as a function of crystal field energy $\Delta$ for the estimated respective values of $J_{ex}$.  The light areas represent regions of induced moment stability.  The red circle marks the measured singlet-doublet transition energy, and the width of the blue hatched region corresponds to the relaxational broadening of the transition.  This illustrates that the induced moment is stable in PrAu$_2$Ge$_2$ because of the strength of the exchange interactions whereas it is unstable in  PrAu$_2$Si$_2$ because of its proximity to the phase boundary.
 \label{Fig5}}
\end{figure*}

If static disorder is not responsible for the frustrated ground state, we must seek an alternative explanation, one that should take into account the observation in Figure 2 that spin glass freezing only occurs in proximity to the induced moment phase boundary.  We propose that relaxational broadening of the singlet-doublet transition provides the key.  Figure 4b shows the temperature dependence of the transition half-width in PrAu$_2$Si$_2$, which increases from about 0.13~meV at 1.6~K to 0.26~meV at 20~K.  The data are consistent with standard models of $f$-electron relaxation due to fluctuations in the single-site crystal field populations, which causes the rapid increase at low temperatures, and conduction electron scattering \cite{Jensen 1987}.  The important result is that the width remains substantial down to the lowest temperature, with the FWHM only falling to a value of 43\% of the transition energy at the glass transition.

This energy broadening will play a significant role in disrupting induced moment formation.  Figure 5 shows the complementary phase diagram to Figure 2, in which $J_{ex}$ is fixed to the estimated values in PrAu$_2$Si$_2$ and PrAu$_2$Ge$_2$ and $T_c$ is plotted against $\Delta$.  The induced moments are predicted to be stable in both compounds, but, if we superimpose the broadened peaks, we see that the peak traverses the phase boundary significantly in the silicon compound but not in the germanium compound.  Strictly speaking, the width is proportional to the inverse lifetime of the excited state and not the lifetime of the induced ground state, but it makes it highly plausible that dynamic fluctuations would disrupt the stability of such moments in PrAu$_2$Si$_2$, but not in PrAu$_2$Ge$_2$.

In induced moment systems, a precursor of long-range order is the appearance in neutron scattering of a quasielastic ``central" peak close to the wavevector of the ordered phase, which is caused by fluctuating regions of short-range magnetic order \cite{Lindgard 1983}. For example, in Pr$_3$Tl, the correlation length associated with this central peak diverges as the critical temperature is approached  \cite{Als-Nielsen 1977}.  Our conjecture is that any divergence of the magnetic correlation length in PrAu$_2$Si$_2$ is suppressed by dynamic fluctuations that limit the lifetime of induced moments and so introduce magnetic site and exchange disorder.  This scenario is similar to the  `avoided criticality' discussed in the theory of structural glass transitions by Tarjus \textit{et al}  \cite{Tarjus 2005}.  In their review, they argue that many glasses are close to a conventional second order phase transition, but that frustration prevents the divergence of the correlation length.  This would make the PrAu$_2$(Si$_{1-x}$Ge$_x$)$_2$ series promising candidates to test their scaling predictions since the degree of frustration can be tuned by varying $J_{ex}$ or $\Delta$ with dopant concentration or pressure.   

There is considerable interest in the ways that spin systems respond to the presence of substantial frustration.  Historically, this subject was stimulated by the observation of spin glass freezing in disordered alloys and compounds, and has more recently focused on systems where the lattice geometry produces a macroscopic degeneracy of possible spin configurations.  We suggest that PrAu$_2$Si$_2$ reveals a new avenue to achieving frustration in systems with neither static disorder nor geometrically frustrated lattices, through dynamic fluctuations, either thermal or quantum, in proximity to a critical phase boundary.

\begin{acknowledgments}
We thank Amir Murani and Ross Stewart for scientific discussions and assistance with the ILL experiments.  This work was supported by the U.S. Department of Energy Office of Science, under Contract No. DE-AC02-06CH11357. 
\end{acknowledgments}

\end{document}